\documentclass[useAMS,usenatbib]{mn2e}
\usepackage{graphicx}
\usepackage{color}
\usepackage{soul}
\usepackage{bm}
\usepackage{amssymb,amsfonts,amsmath}%

\newcommand{\bec}[1]{\mbox{\boldmath $ #1$}}

{}
{}
{}

\newcommand{\meanAAA}{\overline{\mathsf{A}}}

{}
{}
{}
{}
{}
{}
{}
\newcommand{\meanAA}{\overline{\mbox{\boldmath $A$}}{}}{}
\newcommand{\meanBB}{\overline{\mbox{\boldmath $B$}}{}}{}
{}
{}
{}
{}
{}
{}
{}
{}

{}
{}
{}

\newcommand{\meanB}{\overline{B}}

\def\half{{\textstyle{1\over2}}}

\title[The mean tilt of sunspot bipolar regions]
{The mean tilt of sunspot bipolar regions: theory, simulations and comparison with observations}

\author[N. Kleeorin, N. Safiullin, K. Kuzanyan, I. Rogachevskii, A. Tlatov, S. Porshnev]
{ N. Kleeorin,$^{1,2}$
  N. Safiullin,$^{3}$
  K. Kuzanyan,$^{4,5}$
  I. Rogachevskii,$^{1,2}$
  A. Tlatov,$^{6}$
  S. Porshnev$^{3,7}$ \\
 $^{1}$ Department of Mechanical Engineering,
        Ben-Gurion University of Negev, POB 653, 84105 Beer-Sheva, Israel\\
 $^{2}$ Nordita, KTH Royal Institute of Technology and Stockholm University,
        Roslagstullsbacken 23, SE-10691 Stockholm, Sweden\\
 $^{3}$ Institute of Radioelectronics and Information Technology,
        Ural Federal University, 19 Mira str., 620002 Ekaterinburg, Russia\\
 $^{4}$ IZMIRAN, Troitsk, Moscow Region 108840, Russia\\
 $^{5}$ Key Laboratory of Solar Activity, National Astronomical Observatories,
        Chinese Academy of Sciences, 100101 Beijing, China\\
 $^{6}$ Kislovodsk Mountain Astronomical Station of Pulkovo Observatory, Kislovodsk, Russia\\
 $^{7}$ N.N. Krasovskii Institute of Mathematics and Mechanics (IMM UB RAS), Ekaterinburg, Russia}

\begin{document}

%\date{\today}

%\pagerange{\pageref{firstpage}--\pageref{lastpage}}
%\pubyear{2015}

\maketitle

%\label{firstpage}

\begin{abstract}
A theory of the mean tilt of sunspot bipolar regions
(the angle between a line connecting the leading and
following sunspots and the solar equator)
is developed.
A mechanism of formation of the mean tilt is related to the
effect of Coriolis force on meso-scale motions of super-granular convection
and large-scale meridional circulation.
The balance between the Coriolis force and the Lorentz force (the magnetic tension) determines
an additional contribution caused by the large-scale magnetic field
to the mean tilt of the sunspot bipolar regions at low latitudes.
The latitudinal dependence of the solar differential rotation affects the mean tilt
which can explain deviations from the Joy's law for the sunspot
bipolar regions at high latitudes.
The obtained theoretical results and performed numerical simulations
based on the nonlinear mean-field dynamo theory which takes into
account conservation of the total magnetic helicity
and the budget equation for the evolution of the Wolf number density,
are in agreement with observational data of the mean tilt
of sunspot bipolar regions over individual solar cycles 15 - 24.
\end{abstract}


%e-print: NORDITA-2020-003

\maketitle

\section{Introduction}

Origin of solar magnetic field and dynamics of solar activity are the subjects of many studies
and discussions \citep{M78,P79,KR80,ZRS83,RH04,O03,BS05}.
The solar magnetic fields are observed in the form of sunspots and active regions.
One of the characteristics of the solar bipolar region is mean tilt.
The tilt is defined as the angle between
a line connecting the leading and following sunspots and the solar equator.

According to the Joy's law the mean tilt of sunspot bipolar regions
increases with the latitude \citep{H1919,H91,SG99,PB14,MNL14,MN16}.
The mean tilt of sunspot bipolar regions is caused
by effect of the Coriolis force on large-scale motions
in super-granular turbulent convection  \citep{FF00,PB14}.
The Coriolis force is proportional to $\sin \phi$, where $\phi$
is the latitude, so that the main dependence of the mean tilt on the latitude
is expected to be proportional to $\sin \phi$.
The mean tilt of sunspot bipolar regions has been also explained
by the onset of the kink instability \citep{L69,LLP99,HC04}.
The mean tilt can be affected by the large-scale
solar magnetic field \citep{B61,NG05}.

In spite of various theoretical, numerical and observational
studies related to the mean tilt
\citep{DC93,KS08,DS10,MN13,TIS13,ITS15,T15,TVP15},
some observational features related to the mean tilt of
sunspot bipolar regions are not explained.
As follows from observations \citep{TT18}, the latitudinal dependence of the
mean tilt of sunspot bipolar regions can deviate from $\sin \phi$.
There is a non-zero mean tilt of sunspot bipolar regions
at the equator where the Coriolis force vanishes.
In particular, there is a systematic non-zero tilt at the equator
with negative offset for odd cycles and positive offset for even cycles.
\cite{TT18} have also found that the latitudinal dependence of the tilt
varies from one solar cycle to another.
In order to investigate the latitudinal dependence of the mean tilt
of sunspot bipolar regions and its variations in different solar cycles, \cite{TT18}
have used the data of the nearly one century long series
of the magnetic field observations of sunspots from Mount Wilson Observatory.

In the present study we develop a theory of the mean tilt
of sunspot bipolar regions, taking into account the effects of the
solar large-scale magnetic field and the solar differential rotation
on the mean tilt.
We perform the mean-field simulations using the nonlinear mean-field dynamo model
which takes into account conservation of the total magnetic helicity
and the budget equation for the evolution of the Wolf number density.
We have demonstrated that the balance between the Coriolis force and the magnetic tension determines
an additional contribution of the large-scale magnetic field
to the mean tilt of the sunspot bipolar regions at the low latitudes.
We have shown that the latitudinal dependence of the solar differential rotation
affects the mean tilt, explaining the observed deviations from the Joy's
law for the mean tilt for the sunspot bipolar regions at the higher latitudes.
The obtained theoretical and numerical results
have been compared with the latitudinal dependence of the mean tilt
found in observations during the last nine solar cycles.

\section{The theory for the mean tilt of sunspot bipolar regions}

The mean tilt of the sunspot bipolar regions is mainly determined by the
effect of Coriolis force on meso-scale motions of super-granular convection
and large-scale meridional circulation.
We use the momentum, induction and entropy equations applying the anelastic approximation and
neglecting dissipation at the boundary between the solar convective zone and the photosphere:
\begin{eqnarray}
&& {\partial {\bm U} \over \partial t} = - {\bm \nabla}\left({P_{\rm tot} \over \rho_0}\right) - {\bm g}\, S
+ {1 \over 4 \pi \rho_0} ({\bm B} \cdot {\bm \nabla}) {\bm B}
\nonumber\\
&&\quad + {{\bm \Lambda}_\rho \over 8 \pi \rho_0} {\bm B}^2 + {\bm U} \times \left(2 {\bm \Omega}
+ {\bm W}\right)  ,
 \label{B1}
\end{eqnarray}
\begin{eqnarray}
&& {\partial {\bm B} \over \partial t} = ({\bm B} \cdot {\bm \nabla}) {\bm U} - ({\bm U} \cdot {\bm \nabla}) {\bm B} - {\bm B} \, ({\bm \nabla}  \cdot {\bm U})  ,
\label{B2}
\end{eqnarray}
\begin{eqnarray}
&& {\partial S \over \partial t} = - ({\bm U} \cdot {\bm \nabla}) S - {\Omega_b^2 \over g} \, {\bm U} \cdot \hat{\bm r},
 \label{B3}
\end{eqnarray}
\begin{eqnarray}
&& {\bm \nabla}  \cdot {\bm U} =  {\bm \Lambda}_\rho \cdot {\bm U} ,
\label{B4}
\end{eqnarray}
where ${\bm U}$ and ${\bm B}$ are the velocity and magnetic fields,  ${\bm W}={\bm \nabla} \times {\bm U}$ is the vorticity, $P_{\rm tot}=P + {\bm B}^2/ 8 \pi + {\bm U}^2 / 2$ is the total pressure,
$S$ and $P$ are the entropy and pressure of plasma, and
$\Omega_b^2=-({\bm g} \cdot {\bm \nabla}) S_0$.
Here $\rho_0$ and $S_0$ are the plasma density and entropy in the basic reference state,
${\bm \Lambda}_\rho= - {\bm \nabla} \ln \rho_0$, $\, {\bm g}$ is the acceleration due to the gravity, $\hat{\bm r}={\bm r}/|{\bm r}|$ is the unit vector in the radial direction, and ${\bm \Omega}$ is the angular velocity.

\subsection{Effect of the large-scale magnetic field on the mean tilt}

We decompose the solution of equations~(\ref{B1})--(\ref{B4}) as the sum of the equilibrium fields (with superscript ''eq") related to both, the meridional circulation and differential rotation, and perturbations (with tilde) related to both, super-granular motions in the convective zone and rotational motions of sunspots in the photosphere, which contribute to the mean tilt of sunspot bipolar regions,
i.e., ${\bm U}={\bm U}^{\rm eq} +\tilde{\bm u}$, ${\bm B}={\bm B}^{\rm eq}+\tilde{\bm b}$,
$S={S}^{\rm eq}+\tilde{\bm s}$ and $P={P}^{\rm eq}+\tilde{\bm p}$, where
$\tilde{\bm u}= \partial {\bm \xi} / \partial t + {\bm v}^{({\rm c})}$ and ${\bm v}^{({\rm c})}$ is the convective velocity related to the super-granular motions.
The equilibrium magnetic field ${\bm B}^{\rm eq}$ includes the mean magnetic field caused by the dynamo and
the magnetic field of bipolar active regions. The magnetic field of the sunspots is much larger than that
of the mean magnetic field caused by the solar dynamo.
Equations~(\ref{B1})--(\ref{B4}) allow us to obtain an equation for small perturbations ${\bm \xi}$ related to the rotational motions of sunspots at the boundary between the solar convective zone and the photosphere as
\begin{eqnarray}
&& {\partial^2 {\bm \xi} \over \partial t^2} = - {\bm \nabla}\left({\tilde p_{\rm tot} \over \rho_0}\right) - \hat{\bm r} ({\bm \xi} \cdot \hat{\bm r}) \left(\Omega_b^{'2} + \Lambda_\rho^2 \, {U}_{\rm A}^{\,2}\right)
\nonumber\\
 &&\quad + 2 \left({\bm U}^{\rm eq}  + {\bm v}^{({\rm c})} \right) \times {\bm \Omega} + \left({\bm U}_{\rm A} \cdot {\bm \nabla}\right)^2 {\bm \xi}
 \nonumber\\
 &&\quad + \Lambda_\rho \, \left({\bm U}_{\rm A} \cdot {\bm \nabla}\right) \left[ \hat{\bm r} \left({\bm U}_{\rm A} \cdot {\bm \xi}\right) - {\bm U}_{\rm A}({\bm \xi} \cdot \hat{\bm r})\right],
 \label{B5}
\end{eqnarray}
where $\tilde p_{\rm tot}$ are the perturbations of the total pressure,
$\Omega_b^{'2}=\Omega_b^{2} + g (\hat{\bm \xi} \cdot {\bm \nabla}){S}^{\rm eq} / (\hat{\bm \xi} \cdot \hat{\bm r})$, the Alfv\'{e}n speed is ${\rm U}_{\rm A}={\rm B}^{\rm eq}/ (4 \pi \rho_0)^{1/2}$ and $\hat{\bm \xi}={\bm \xi}/|{\bm \xi}|$ is the unit vector.
To derive equation~(\ref{B5}), we rewrite equations~(\ref{B2}) and~(\ref{B3}) for small perturbations $\tilde{\bm b}$ and $\tilde{s}$ [see equations~(\ref{BBB2}) and~(\ref{BBB3}) in Appendix~A], and substitute them to equation~(\ref{B1})
rewritten for small perturbations ${\bm \xi}$. We also assume that
\begin{eqnarray}
&& |\partial {\bm \xi} / \partial t| \ll |{\bm v}^{({\rm c})}|, \quad |\partial {\bm \xi} / \partial t| \ll |{\bm U}^{\rm eq}| ,
\nonumber\\
&& \Omega \ll \tau_{_{\rm A}}^{-1},  \quad \Omega \ll \tau_c^{-1} ,
\nonumber\\
&& \tilde L_B \gg H_\rho,  \quad \tilde L_B \gg L_\xi,
\label{AB5}
\end{eqnarray}
where $\tau_{_{\rm A}} = L_B /{U}_{\rm A}$ is the maximum Alfv\'{e}n time,
$L_B$ is the length of the magnetic field line,
$\tau_{c}=H_\rho/v_r^{({\rm c})}$ is the characteristic time associated
with convective super-granular motions,
$\tilde L_B$ is the characteristic spatial scale of the magnetic field ${\rm B}^{\rm eq}$ variations,
$L_\xi$ is the characteristic scale of variations of ${\bm \xi}$ and $H_\rho=|{\bm \nabla} \ln \rho_0|^{-1}$
is the density stratification hight.
We also consider an equilibrium in the absence of rotation.

Let us discuss the physical meaning of different terms in equation~(\ref{B5}). The term $\hat{\bm r} ({\bm \xi} \cdot \hat{\bm r}) \Omega_b^{'2}$ describes the internal gravity waves, while the term $\hat{\bm r} ({\bm \xi} \cdot \hat{\bm r}) \Lambda_\rho^2 \, {U}_{\rm A}^{\,2}$ contributes to the slow magneto-acoustic waves. The term, $\left({\bm U}_{\rm A} \cdot {\bm \nabla}\right)^2 {\bm \xi}$ describes the
Alfv\'{e}n waves, and the last two terms ($\propto \Lambda_\rho$) in equation~(\ref{B5}) are the magnetic tension in the density stratified medium which contribute to the fast magneto-acoustic waves. Other terms are the Coriolis force and the gradient of the total pressure.

We define the tilt of the sunspot bipolar regions using the vector ${\bm \delta}^{\rm tw} = {\bm \nabla} \times {\bm \xi}$, which is related to the perturbations of vorticity, $\tilde{\bm w} \equiv (\partial / \partial t) ({\bm \nabla} \times {\bm \xi}) \equiv \partial {\bm \delta}^{\rm tw} / \partial t$.
The absolute value $|{\bm \delta}^{\rm tw}| \approx |\tilde{\bm w}| \, \delta t$ of this vector characterises the twist of the magnetic field lines which connect the sunspots of the opposite magnetic polarity
of the bipolar region. During the twist time $\delta t$, the perturbations of the vorticity $\tilde{\bm w}$ do not vanish. The direction of the vector ${\bm \delta}^{\rm tw}$ coincides with that of the vorticity $\tilde{\bm w}$, and it is perpendicular to the plane of the twist. Therefore, the radial component of the vector ${\bm \delta}^{\rm tw}$ at the boundary between the convective zone and the photosphere can characterise the tilt of the sunspot bipolar regions. At this boundary the magnetic field inside the sunspots is preferably directed in the radial direction.
The mean tilt $\gamma \equiv \langle{\bm \delta}^{\rm tw} \cdot {\bm e}_B\rangle_{\rm time}$ of sunspot bipolar regions at the surface of the Sun is defined by averaging of the scalar product ${\bm \delta}^{\rm tw} \cdot {\bm e}_B$ over the time that is larger than the maximum Alfv\'{e}n time $\tau_{_{\rm A}}$,
where ${\bm e}_B={\bm B}^{\rm eq} / {B}^{\rm eq}$ is the unit vector along the large-scale magnetic field ${\bm B}^{\rm eq}$.

The details of the derivation of the equation for the mean tilt at the solar surface $\gamma$ are given in Appendix~\ref{Appendix-A}. Here we give the derived expression for the mean tilt of the sunspot bipolar regions at the surface as
\begin{eqnarray}
\gamma = {\tau_{_{\rm A}}^2 \over 2 \pi} \Big\langle{\bm \nabla} \times \left(\left({\bm U}^{\rm eq} + {\bm v}^{({\rm c})} \right) \times {\bm \Omega}\right)\Big\rangle_{\rm time} \cdot {\bm e}_B ,
 \label{B14}
\end{eqnarray}
where the angular brackets $\langle ... \rangle_{\rm time}$ denote the averaging over the time that is larger than the maximum Alfv\'{e}n time $\tau_{_{\rm A}}$.
We also assume that the source of the tilt of the sunspot bipolar regions, $I_\gamma=2 \Big[{\bm \nabla} \times [({\bm U}^{\rm eq} + {\bm v}^{({\rm c})}) \times {\bm \Omega}]\Big] \cdot {\bm e}_B$, is localized
at the vicinity of the boundary between the solar convective zone and the photosphere.
Calculating the source $I_\gamma$ and averaging it over the time larger than
the maximum Alfv\'{e}n time, we arrive at the expression for the mean tilt of sunspot bipolar regions as
\begin{eqnarray}
\gamma &=& - \delta_0
\biggl[\sin \phi - \cos \phi  \, {\tau_{c} \over R_\odot} \, {\partial \overline{U}_r \over \partial \phi}
\biggr] ,
 \label{B15}
\end{eqnarray}
where $\delta_0 = (1 + \tilde C) \, \tau_{_{\rm A}}^2 \, \Omega / (2 \pi \, \tau_{c})$, $R_\odot$ is the solar radius, and $\phi$ is the latitude. Here we also took into account that $\partial v_r^{({\rm c})} / \partial r \approx - \tilde C \, v_r^{({\rm c})} / H_\rho$ and $\langle \partial v_r^{({\rm c})} /\partial \phi\rangle =0$.
The radial mean velocity, $\overline{U}_r$ is estimated as
\begin{eqnarray}
\overline{U}_r = {C_u  \over \kappa} \, \left({\ell_{\rm top}^2 \over R_\odot} \right) \, \left({\rho_{\rm bot} \over \rho_{\rm top}} \right)\, \left({u_{\rm bot}^2 \over \nu_{_{T}}^{\rm top}} \right) \,  \left({\partial^2  \over \partial \phi^2}\, {\overline{\bm B}^2 \over \meanB_{\rm eqp}^2} \right)_{\rm bot} ,
 \label{R7}
\end{eqnarray}
(see Appendix~\ref{Appendix-B}), where
$\ell_{\rm top}$ is the integral scale of turbulent motions in the upper part of the convective zone, $\rho_{\rm bot}$ and $\rho_{\rm top}$ are the plasma densities in the bottom and upper parts of the convective zone, respectively, $u_{\rm bot}$ and $\nu_{_{T}}^{\rm top}$ are the characteristic turbulent velocity and the turbulent viscosity, respectively, in the upper part of the convective zone, and
$\meanB_{\rm eqp} = u \, \sqrt{4 \pi \rho}$ is the equipartition magnetic field.
The parameter $\kappa \approx 0.3$ -- $0.4$ characterises
a fraction of the large-scale radial momentum of plasma  which is lost
during crossing the boundary between the convective zone and photosphere.
The constant $C_u$ in equation~(\ref{R7}) varies from $0.7$ to $1$
depending on the radial profile of the mean magnetic field.
Substituting equation~(\ref{R7}) in equation~(\ref{B15}),
we obtain the expression for the mean tilt of the sunspot bipolar regions as
\begin{eqnarray}
\gamma &=& - \delta_0 \Big[\sin \phi - \delta_{_{M}} \, \cos \phi\Big] ,
 \label{B16}
\end{eqnarray}
where
\begin{eqnarray}
\delta_{_{M}} &=& C_{_{M}} \, \left({\ell_{\rm top} \over R_\odot} \right)^2 \,
\left({\rho_{\rm bot} \over \rho_{\rm top}} \right)\,
\left({\eta_{_{T}}^{\rm bot} \over \eta_{_{T}}^{\rm top}} \right)
\, \left({\tau_{c} \over \tau_{\rm bot}} \right)
\nonumber\\
&& \quad \times \biggl({\partial^3  \over \partial \phi^3}\, {\overline{\bm B}^2 \over \meanB_{\rm eqp}^2} \biggr)_{\rm bot} ,
 \label{B30}
\end{eqnarray}
and $\tau_{\rm bot} =\ell_{\rm bot}/ u_{\rm bot}$ is the characteristic turbulent time at the bottom of the convective zone, $C_{_{M}} = 3 C_u  / (\kappa \, {\rm Pr}_{_{T}}) \approx 10$, ${\rm Pr}_{_{T}}= \nu_{_{T}}/\eta_{_{T}}$ is the turbulent Prandtl number and $\eta_{_{T}}$ is the turbulent magnetic diffusion coefficient.

The parameter $\delta_{_{M}}$ describes the magnetic contribution to the mean tilt of the sunspot bipolar regions.
The mechanism related to the magnetic contribution to the mean tilt is as follows.
The Coriolis force results in the twist of sunspots in the photosphere, and the balance between the Coriolis force and the magnetic tension determines the magnetic contribution $\delta_{_{M}}$ to the mean tilt of the sunspot bipolar regions.
The magnetic contribution $\delta_{_{M}}$ to the mean tilt is important in the vicinity of the equator where the main contribution caused by the Coriolis force $\propto \sin \phi$ vanishes.
Note also that since $\delta_{_{M}} \propto [(\partial^3  / \partial \phi^3)\, (\overline{\bm B}^2
/ \meanB_{\rm eqp}^2)]_{\rm bot}$, the combination of the dipole and quadrupole dynamo modes has a non-zero contribution to $\delta_M$ in the vicinity of the equator.

To estimate the mean tilt of the sunspot bipolar regions,
we use the values of governing parameters taken
from models of the solar convective zone (see, e.g., \cite{BT66,S74}; more modern treatments make little
difference to these estimates). In particular, at depth $ H \sim 2
\times 10^{10}$ cm (i.e., at the bottom of the convective zone),
the magnetic Reynolds number ${\rm Rm}^{\rm bot} = u_{\rm bot}
\, \ell_{\rm bot} /\eta = 2 \cdot 10^9$ (where $\eta$ is the magnetic diffusion
coefficient due to electrical conductivity of plasma), the turbulent velocity
$u_{\rm bot} \sim 2 \times 10^3 $ cm s$^{-1}$, the integral scale of turbulence
$\ell_{\rm bot} = 8 \times 10^9$ cm, the plasma density $ \rho_{\rm bot} =
2 \times 10^{-1}$ g cm$^{-3}$, and the turbulent diffusion coefficient
$\eta_{_{T}}^{\rm bot} = 5.3 \times 10^{12}$
cm$^2$s$^{-1}$. The density stratification scale is estimated here
as $H_\rho^{\rm bot} = \rho / |\nabla \rho| = 6.5 \times 10^9$ cm
and the equipartition mean magnetic field $\meanB_{\rm eqp}^{\, {\rm bot}} = 3000$ G. In
the upper part of the convective zone, say at depth $H \sim 2
\times 10^7$ cm, these parameters are
${\rm Rm}^{\rm top} = u_{\rm top} \, \ell_{\rm top} /\eta = 10^5$,
$u_{\rm top} = 9.4 \times 10^4 $ cm s$^{-1}$, $\ell_{\rm top} = 2.6 \times 10^7$
cm, $\rho_{\rm top} = 4.5 \times 10^{-7}$ g cm$^{-3}$, $\eta_{_{T}}^{\rm top} =
0.8 \times 10^{12} $ cm$^2$ s$^{-1}$, $H_\rho^{\rm top} = 3.6
\times 10^7$ cm, and the equipartition mean magnetic field is $\meanB_{\rm eqp}^{\, {\rm top}}
= 220 $ G.

Using these estimates, we calculate the parameters $\delta_0$ and $\delta_{_{M}}$ which determine
the mean tilt of the sunspot bipolar region. Taking the Alfv\'{e}n speed
${U}_{\rm A}= 5 \times 10^4$ cm s$^{-1}$,
the length the magnetic field line $L_B=4 H_\rho = 4 \times 10^9$ cm, we obtain the  Alfv\'{e}n time
$\tau_{_{\rm A}} = L_B /{U}_{\rm A} = 10^5$ s. Taking the convective velocity
$u_c = (3$ -- $5) \times 10^4 $ cm s$^{-1}$, we obtain the convective time as $\tau_c=(2$ -- $3) \times 10^4$ s.
This yields $\delta_0= 0.3$ -- $0.5$ (in radians) and $\delta_{_{M}} = 0.05$ -- $0.2$.
This implies that the magnetic contribution  $\delta_{_{M}}$ to the mean tilt $\gamma$ is essential only in the low latitude region where $\sin \phi$ is small.

The main uncertainty for the estimate of the parameter $\delta_{_{M}}$ is related to the estimate of the third
derivative of the mean magnetic field with respect to the latitude [see equation~(\ref{B30})].
This is the reason why we use the numerical dynamo model for the estimation of this quantity (see
Section~3).
The additional uncertainty is related to the parameters of turbulence at the bottom of
the solar convective zone, where the Coriolis number ${\rm Co} \equiv 2 \Omega \tau \gg 1$,
where $\tau$ is the characteristic turbulent time.
This effect has not been taken into account in the standard
models of the solar convective zone based on the mixing length theory.

\subsection{Effect of latitudinal dependence of the solar rotation on the mean tilt}

In this section we take into account an effect of latitudinal dependence of
the solar differential rotation on the tilt of the sunspot bipolar regions.
In particular, the latitudinal dependence of the solar rotation at the surface of the sun
can be approximated by
\begin{eqnarray}
\Omega=\Omega_0 \left(1 - C_2 \sin^2 \phi - C_4 \sin^4 \phi\right) ,
\label{RQ1}
\end{eqnarray}
[see \cite{LH82}], where $\Omega_0=2.83 \times 10^{-6}$ s$^{-1}$, $C_2=0.121$ and $C_4=0.173$.
Substituting equation~(\ref{RQ1}) into equation~(\ref{B16})
with $\delta_0 = (1 + \tilde C) \, \tau_{_{\rm A}}^2 \, \Omega / (2 \pi \, \tau_{c})$,
we obtain
\begin{eqnarray}
\gamma &=& - \tilde \delta_0 \Big[\sin \phi + \delta_3 \sin 3\phi - \delta_5 \sin 5\phi
\nonumber\\
&&- \tilde \delta_{_{M}} \, \Big(\cos \phi + \tilde \delta_3 \, \cos 3\phi - \tilde \delta_5 \, \cos 5\phi\Big) \Big] ,
 \label{RQ2}
\end{eqnarray}
where $\tilde \delta_0 = C_D \, \delta_0$, $\tilde \delta_{_{M}} = \delta_{_{M}} \tilde C_D / 16 C_D$,
$C_D = 1 - (3 C_2 + 5 C_4)/4 \approx 0.693$, $\tilde C_D = 1 -4C_2 -2 C_4\approx 0.17$ and
$\delta_3 =  (4 C_2 + 5 C_4)/ 16 C_D \approx 0.122$,
$\delta_5 =  C_4/ 16 C_D \approx 1.56 \times 10^{-2}$,
$\tilde \delta_3 =  (4 C_2 + 3 C_4)/ \tilde C_D \approx 4.48$,
and $\tilde \delta_5 = C_4/ \tilde C_D \approx 1.02$.
For the derivation of equation~(\ref{RQ2}) we take into account that $\Omega / H_\rho \gg |\partial \Omega / \partial r|$ and $\Omega / H_\rho \gg r^{-1}|\partial \Omega / \partial \theta|$, and we also use identities~(\ref{RP1})--(\ref{RP4}) given in Appendix~\ref{Appendix-A}.

\begin{figure}
\centering
\includegraphics[width=8.5cm]{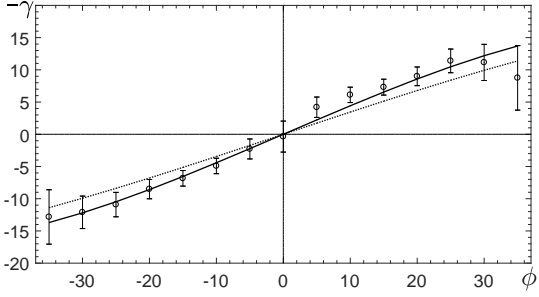}
\caption{\label{Fig1}
The mean tilt $-\gamma$ (in degrees)
versus the latitude $\phi$ (in degrees): Eq.~(\ref{RQ2}) of our theory with
with $\tilde \delta_0 =0.35$, $\delta_3 = 0.12$, $\delta_5 = 1.6 \times 10^{-2}$,
$\tilde \delta_3 = 4.48$, $\tilde \delta_5 = 1.02$ and $\tilde \delta_{_{M}}=0$
(solid line) and the data from observations (circles) of all solar cycles
published in Fig.~3 of Tlatova et al. (2018).
Dotted line corresponds to Eq.~(\ref{B16}) with $\delta_0 =0.406$ and
$\delta_{_{M}}=0$.
}
\end{figure}

\begin{figure}
\centering
\includegraphics[width=8.5cm]{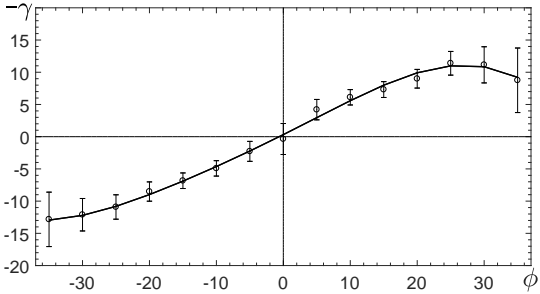}
\caption{\label{Fig2}
The mean tilt $-\gamma$ (in degrees)
versus latitude $\phi$ (in degrees): Eq.~(\ref{RQ2}) of our theory
with $\tilde \delta_0 =0.31$, $\delta_3 = 1.26$, $\delta_5 = 0.22$,
$\tilde \delta_3 = 1.87$, $\tilde \delta_5 = 0.78$ and $\tilde \delta_{_{M}}=0.2$
(solid line),
and the data from observations (circles) of all solar cycles published in Fig.~3
of Tlatova et al. (2018).
}
\end{figure}

In Figure~\ref{Fig1} we show the mean tilt $-\gamma$ (solid line)
versus latitude $\phi$ given by equation~(\ref{RQ2}) of our theory, where $\gamma$ and $\phi$
are measured in degrees, and
we use the following values of parameters:  $\tilde \delta_0 =0.35$, $\delta_3 = 0.12$, $\delta_5 = 1.6 \times 10^{-2}$, $\tilde \delta_3 = 4.48$, $\tilde \delta_5 = 1.02$ and $\tilde \delta_{_{M}}=0$ (i.e., the magnetic contribution to the mean tilt of the sunspot bipolar regions is neglected here).
For comparison with observations, we also show in Figure~\ref{Fig1} the data obtained from observations of all solar cycles presented in Figure~3 of \cite{TT18} and shown here as circles (see Section~4 for more details
about the observational data).
The observational data have been averaged over bipolar regions of all sizes [see Tables~1 and~2 in \cite{TT18}], where the mean value and the standard deviation of Gaussian fittings have been computed.
Dotted line in Figure~\ref{Fig1} corresponds to Eq.~(\ref{B16})
which does not taken into account the effect of the latitudinal dependence of the solar rotation on the mean tilt. Figure~\ref{Fig1} demonstrates that the account for the latitudinal part of the differential rotation improves the agreement with observations.

In Figure~\ref{Fig2} we also show the theoretical latitudinal dependence of the mean tilt $-\gamma$ (solid line),
taking into account the magnetic contribution to the mean tilt
of the sunspot bipolar regions ($\tilde \delta_{_{M}}=0.2$).
Slightly varying the values of other coefficients ($\tilde \delta_0 =0.31$, $\delta_3 = 1.26$, $\delta_5 = 0.22$,
$\tilde \delta_3 = 1.87$ and $\tilde \delta_5 = 0.78$), we obtain a good agreement between
the theoretical predictions for the mean tilt and the observational data (see Figure~\ref{Fig2}).

\section{Numerical modelling of the mean tilt of sunspot bipolar regions}

To obtain the time evolution of the mean tilt of sunspot bipolar regions, in particular, to get the butterfly diagram of the mean tilt, we use a nonlinear mean-field dynamo model discussed in details by \cite{KSR16,SKR18}. Below we briefly outline this model.
We use spherical coordinates $(r, \theta, \varphi)$ for an
axisymmetric large-scale magnetic field, $ \meanBB = \meanB_\varphi
\bec{e}_{\varphi} + \bec{\nabla} {\bf \times} (\meanAAA \bec{e}_{\varphi})$.
We consider the mean-field dynamo equations in a thin convective shell,
where we take into account strong variation of the plasma density
in the radial direction by averaging the dynamo equations for the
mean toroidal field $\meanB_\varphi$ and the magnetic potential $\meanAAA$
of the mean poloidal field over the depth of the convective shell
(so called the no-$r$ dynamo model).
We neglect the curvature of the convective shell
and replace it by a flat slab (see below).
The mean-field dynamo equations for $\meanB_\varphi$ and $\meanAAA$ read
\begin{eqnarray}
{\partial \meanB_\varphi \over \partial t} &=& G D \sin \theta {\partial
\meanAAA \over \partial \theta} + {\partial ^2 \meanB_\varphi \over \partial
\theta^2} - \mu ^2 \meanB_\varphi ,
\label{TM1}
\end{eqnarray}
\begin{eqnarray}
{\partial \meanAAA \over \partial t} &=& \alpha \meanB_\varphi + {\partial^2 \meanAAA
\over \partial \theta^2} - \mu^2 \meanAAA .
\label{TM2}
\end{eqnarray}
In the framework of the no-$r$ model, the last terms in the right hand side of equations~(\ref{TM1}) and~(\ref{TM2}), which determine turbulent diffusion of the mean magnetic field
in the radial direction, are given as $-\mu^2 \meanB_\varphi$
and $-\mu^2\meanAAA$ \citep{KSR16,SKR18}.
The differential rotation is characterised by parameter
$G =\partial \Omega / \partial r$, and the parameter $\mu$ is determined by the following equation:
$\int_{2/3}^{1} (\partial^2 \meanB_\varphi / \partial r^2) \,dr = - (\mu^2/3) \meanB_\varphi$.
The dynamo number $D$ in equation~(\ref{TM1}) is defined as $D = R_\alpha \, R_\omega$, where
$R_{\alpha} = \alpha_0 R_\odot / \eta_{_{T}}$ and $R_\omega = (\delta \Omega) \, R_\odot^2 / \eta_{_{T}}$.
Here the angular velocity $\delta\Omega$
characterises the differential rotation and $\alpha_0$ is the characteristic value
of the kinetic part of the $\alpha$ effect.
When the dynamo number is negative, equations~(\ref{TM1}) and~(\ref{TM2})
describe the dynamo waves propagating from the central
latitudes towards the equator.
We use the standard latitudinal profile of the kinetic part of the $\alpha$ effect as
$\alpha(\theta)=\alpha_0 \sin^3 \theta \cos \theta$.

Equations~(\ref{TM1})--(\ref{TM2}) are written in a non-dimensional form, where
the length is measured in units of the solar radius $R_\odot$, time is measured
in units of the turbulent
magnetic diffusion time $R_\odot^2 / \eta_{_{T}}$, the angular velocity $\delta\Omega$
is measured in units of the maximal value
of $\Omega$, and $\alpha$ is measured in units of the maximum value of the
kinetic part of the $ \alpha $-effect.
Here $\eta_{_{T}}=\ell \, u / 3$ is the turbulent magnetic
diffusion coefficient, where the integral scale of the turbulent motions
$\ell$ and turbulent velocity $u$ at the scale $\ell$ are measured in units of their
maximum values through the convective region, and
the magnetic Reynolds number ${\rm Rm}=\ell \, u/\eta$
is defined using the maximal values of the integral scale $\ell$ and
the characteristic turbulent velocity $u$.
The toroidal component of the mean magnetic field $\meanB_\varphi$ is measured
in the units of the equipartition field $\meanB_{\rm eqp} = u \sqrt{4 \pi \rho_{\rm bot}}$,
and the vector potential $\meanAAA$ of the poloidal component of the mean magnetic field is measured in units of $R_{\alpha} R_\odot \meanB_{\rm eqp}$.
The density $\rho_0$ is normalized by its value $\rho_{\rm bot}$
at the bottom of the convective zone.
The radius $r$ varies from $2/3$ to $1$
inside the convective shell, so that
the value $\mu=3$ corresponds to a convective
zone with a thickness of about 1/3 of the radius.

Let us discuss the main nonlinear effects in the dynamo model.
The total $\alpha$ effect is the sum of the kinetic
and magnetic parts,
$\alpha = \chi_{\rm v} \Phi_{\rm v}(\meanB) + \sigma \chi_{\rm c} \Phi_{\rm m}(\meanB)$
\citep{KSR16,SKR18},
where $\chi_{\rm v} = - (\tau /3) \, \overline{\bec{u}\cdot(\bec{\nabla}
{\bf \times} \bec{u})}$ and $\chi_{\rm c} = (\tau / 12 \pi \rho)\, \overline{{\bm b} \cdot (\bec{\nabla} {\bf \times} {\bm b})}$. Here $\tau$ is the correlation time of the turbulent velocity field,
${\bm u}$ and ${\bm b}$ are the velocity and magnetic fluctuations, respectively,
and $\sigma = \int_{2/3}^{1} (\rho_0(r)/ \rho_{\rm bot})^{-1} \, dr$.

The magnetic part of the $\alpha$ effect \citep{FPL75,PFL76} and density of the magnetic helicity are related to the density of the current helicity $\overline{{\bm b} \cdot (\bec{\nabla} {\bf \times} {\bm b})}$  in the approximation of weakly inhomogeneous turbulent convection \citep{KR99}.
The quenching functions $\Phi_{\rm v}(\meanB)$ and $\Phi_{\rm m}(\meanB)$ in
equation for the total $\alpha$ effect are given by:
$\Phi_{\rm v}(\meanB) = (1/7) [4 \Phi_{\rm m}(\meanB) + 3 \Phi_{\rm B}(\meanB)]$ and
$\Phi_{\rm m}(\meanB) = (3/ 8\meanB^2) \, [1 - \arctan (\sqrt{8} \meanB) /
\sqrt{8} \meanB]$  \citep{RK00,RK01,RK04},
where $ \Phi_{\rm B}(\meanB) = 1 - 16 \meanB^{2} + 128 \meanB^{4} \ln [1 + 1/(8\meanB^2)]$,
and $\chi_{\rm v}$ and $\chi_{\rm c}$ are measured in units of maximal
value of the $\alpha$-effect.
The function $\Phi_{\rm v}$ describes the algebraic quenching of the
kinetic part of the $\alpha$ effect that
is caused by the feedback effects of the mean magnetic field
on the turbulent electromotive force.
The densities of the helicities and quenching functions are
associated with a middle part of the convective zone.
The parameter $\sigma > 1$ is a free parameter.

The magnetic part $\alpha_{\rm m}$ of the $\alpha$ effect is based on two
nonlinearities: the algebraic nonlinearity (quenching of $\alpha_{\rm m}$), given by the function
$\Phi_{\rm m}(\meanB)$, and the dynamic nonlinearity.
In particular, the function $\chi_{\rm c}(\meanBB)$ is determined by a dynamical
equation \citep{KR82,KR99,KRR95,KMR00,KMR02,KMR03,KMRS03,BS05,ZKR06,ZKR12}:
\begin{eqnarray}
&& {\partial \chi_{\rm c} \over \partial t} + \left(\tau_\chi^{-1} + \kappa_{_{T}}
\mu^2\right)\chi_{\rm c} = 2\left({\partial
\meanAAA \over \partial \theta} {\partial \meanB_\varphi \over \partial \theta} + \mu^2 \meanAAA \, \meanB_\varphi\right)
\nonumber\\
&& \quad- \left({R_\odot^2 \over 2 \ell^2}\right) \, \alpha \, \meanB^2 - {\partial \over \partial \theta} \left(\meanB_\varphi {\partial \meanAAA \over \partial \theta} - \kappa_{_{T}} {\partial \chi^c
\over \partial \theta} \right) ,
\label{TM9}
\end{eqnarray}
where ${\bm F}_\chi = -\kappa_{_{T}} \bec{\nabla} \chi_{\rm c}$
is the turbulent diffusion flux of the density of the magnetic helicity,
$\kappa_{_{T}}$ is the coefficient of the turbulent diffusion of the magnetic helicity,
$\tau_\chi = \ell^2 / \eta$ is the relaxation time of magnetic helicity.
This dynamical equation is derived from the conservation law for the total
magnetic helicity.
The inverse time $\tau_\chi^{-1}$ averaged over the depth of the convective zone
is given by
\begin{eqnarray}
\tau_\chi^{-1} = H^{-1} \int_{2/3}^{1} \tilde \tau_\chi^{-1}(r) \,d r \sim {H_\ell \, R_\odot^2 \,
\eta \over H \, \ell^2 \, \eta_{_{T}}} ,
\label{TM11}
\end{eqnarray}
where $H$ is the depth of the convective zone,
$H_\ell$ is the characteristic scale of variations of the integral turbulence scale $\ell$, and
$\tilde \tau_\chi(r) = (\eta_{_{T}} / R_\odot^{2}) (\ell^2 / \eta)$ is the non-dimensional
relaxation time of the density of the magnetic helicity. The values
$H_\ell , \, \eta, \, \ell$ in equation~(\ref{TM11}) are associated
with the upper part of the convective zone.
The squared mean magnetic field is given by
\begin{eqnarray}
\meanB^2 = {2 \ell^2 \over R_\odot^2} \, \left[\meanB_\varphi^2 + R_\alpha^2 \left(\mu^2 \meanAAA^2 + \left( {\partial \meanAAA \over \partial \theta}\right)^2\right) \right] .
\label{TM10}
\end{eqnarray}

Let us discuss the assumptions we use in the mean-field
dynamo model which we apply for the numerical mean-field simulations.
In the used dynamo model, equations are averaged over the depth of the solar convective zone in the radial direction. Such averaging is made because the fluid density in the solar convective zone is stratified by seven orders of magnitude. There is no any numerical dynamo model which is able to take into account such strong fluid density stratification in the radial direction.
The reason is that the numerical simulations should have very high spatial resolution to resolve the convective zone with such strong density stratification, which is not real.
That is why we use the no-$r$ dynamo model.

From available observations, there is no any information
about the radial profile of the kinetic helicity and the alpha effect
in the convective zone of the Sun.
This implies that numerical mean-field dynamo models are based on the
assumption about the radial profile of the alpha effect,
which causes an uncertainty in the radial profile of the numerical solutions.

On the other hand, three-dimensional mean-field dynamo models
allow to obtain non-axisymmetric dynamo modes
and to study non-axisymmetric effects, e.g., solar active longitudes
\citep{BMS06,BR04,PK15}.
In particular, observations show that solar activity is distributed non-axisymmetrically,
concentrating at ''preferred longitudes."
This effect appears when the solar activity persists
within a fixed interval of longitudes for a long period of time.

Note also that radial dependencies of the $\alpha$ effect and differential rotation
may give new features.
For example, the change of the sign of the $\alpha$-effect either with radius or latitude can give a poleward branch of the solar activity \citep{Y81,G85,K98}.
Furthermore, there are indirect signatures that the sign of the observable current helicity, the proxy of the $\alpha$-effect, change with depth in the solar convection zone \citep{KL03}.
Similarly, to obtain simultaneously coexisting poleward and equatorward branches of the dynamo waves, a two-dimensional dynamo model with different signs of the differential rotation can be considered \citep{BKS00}.

The no-curvature assumption is used in the dynamo model to take into account the polar regions,
where the exact calculations of the Stokes operator require very high resolution.
On the other hand, we use the mean-field numerical simulations
only for the calculations of the third-order derivative of the mean magnetic
field with respect to the latitude which is needed to determine
the time evolution of the magnetic contribution $\delta_{_{M}}$ to the mean tilt.

The observed solar activity is characterised by
the Wolf number \citep{G73,S89}, defined
as $W= 10 g_{\rm w} +f_{\rm w}$, where $g_{\rm w}$ is the number of sunspot groups
and $f_{\rm w}$ is the total number of sunspots in the visible part of the sun.
The dynamo model applied in the present study,
is directly related to the evolution of the Wolf number.
In particular, we derive the phenomenological budget equation for the surface density
of the Wolf number \citep{KSR16,SKR18}, that is given
in Appendix~\ref{Appendix-C} [see equation~(\ref{TM12})].
This equation allows us to perform direct comparisons between
the numerical solution of the dynamo equations and
the observational data for the evolution of the Wolf number.
The used budget equation for the surface density
of the Wolf number contains the source term for the sunspot formation
(i.e., the rate of production of the Wolf number density)
and the sink term describing the decay of sunspots.
The rate of production of the Wolf number density depends on two control parameters:
the threshold $\overline{B}_{\rm cr}$ in the mean magnetic field
required for the formation of sunspots, and the inverse time $\gamma_{\rm inst}$ of
the formation of sunspots. The form of the budget equation for
evolution of the Wolf number is rather general.

As an example for estimation of the parameters $\overline{B}_{\rm cr}$ and $\gamma_{\rm inst}$,
we use the negative effective magnetic pressure instability
which can be excited even for uniform mean magnetic field.
This effect has been investigated in analytical
\citep{KRR89,KRR90,KMR93,KMR96,KR94,RK07}
and numerical studies \citep{BKR11,BRK16}. This instability
results in formation of magnetic spots \citep{BKR13,BGKR14}
and bipolar active regions \citep{WKR13,WKR16}.

There is also another mechanism for the
formation of the large-scale inhomogeneous magnetic structures, e.g., the magnetic
buoyancy instability of stratified continuous magnetic field
\citep{P66,P79,G70a,P82}.
This instability is excited when the characteristic scale of the initial mean magnetic field
variations is less than the density stratification hight scale.
The critical magnetic field $\overline{B}_{\rm cr}$ and the growth rate
$\gamma_{\rm inst}$ for the magnetic buoyancy instability can be used for the estimation
of the rate of production of the Wolf number density.
For more discussion, see Appendix~C.

The observed Wolf number time series (the monthly mean total sunspot number) have been used
for comparison with the obtained results of the mean-field  numerical simulations.
This observational data are available in open access from the World Data Center SILSO,
Royal Observatory of Belgium.
The details of the quantitative comparisons between the numerical results
and observational data are given by \cite{KSR16,SKR18} and outlined below.

In the present study, we solve numerically equations~(\ref{TM1}), (\ref{TM2}), (\ref{TM9}) and~(\ref{TM12})
for the following initial conditions: $\meanB_\phi(t=0,\theta)=S_1 \sin\theta + S_2 \sin(2\theta)$ and $\meanAAA(t=0,\theta)=0$.
The parameters of the numerical simulations are as follows: $D=-8450$, $G=1$, $\sigma=3$, $\mu=3$, $\kappa_{_{T}}=0.1$, $R_\alpha=2$, $\tau_\chi=6.3$, $S_1=0.051$ and $S_2=0.95$.
The choice of these parameters in the numerical simulations is caused by the following reasons.
In our previous studies \citep{KSR16,SKR18}
we performed a parameter scan using about $10^3$ runs with different sets of parameters
to find an optimal set of parameters to reach a large correlation between
the Wolf numbers obtained in the numerical simulations and observations.
There are two crucial parameters which strongly affect the dynamics
of the nonlinear dynamo system: the dynamo number $D$
and the initial field $B_{\rm init}^{\rm dip}$ for the dipole mode,
determined by the parameter $S_2$.
A proper choice of the initial field $B_{\rm init}^{\rm dip}$ allows
to avoid very long transient regimes.

To find the maximum correlation between the Wolf numbers obtained
in the numerical simulations and observations, the following
parameter scan has been performed: $- 8800 \leq D \leq - 8200$
and $0.85 \leq S_2 \leq 0.95$.
The maximum correlation (with about a 70 \% correlation in observed data
and numerical simulations of Wolf numbers) is obtained when the
parameters are $D=-8450$ and $S_2=0.95$
[see Fig.~12 in \cite{KSR16}].
The parameter $\mu$ determines the critical dynamo number, $|D_{\rm cr}|$,
for the excitation of the large-scale dynamo instability.
The flux of the magnetic helicity [see Eq.~(\ref{TM9})],
characterised by the parameter $\kappa_{_{T}}$, cannot be very small to
avoid the catastrophic quenching of the $\alpha$ effect \citep{KMR00,KMR02,KMR03,KMRS03}.
The optimal value for this parameter is $\kappa_{_{T}} \approx 0.1$.
The variations of the other parameters only weakly affect
the obtained results \citep{KSR16}.

\begin{figure}
\centering
\includegraphics[width=8.5cm]{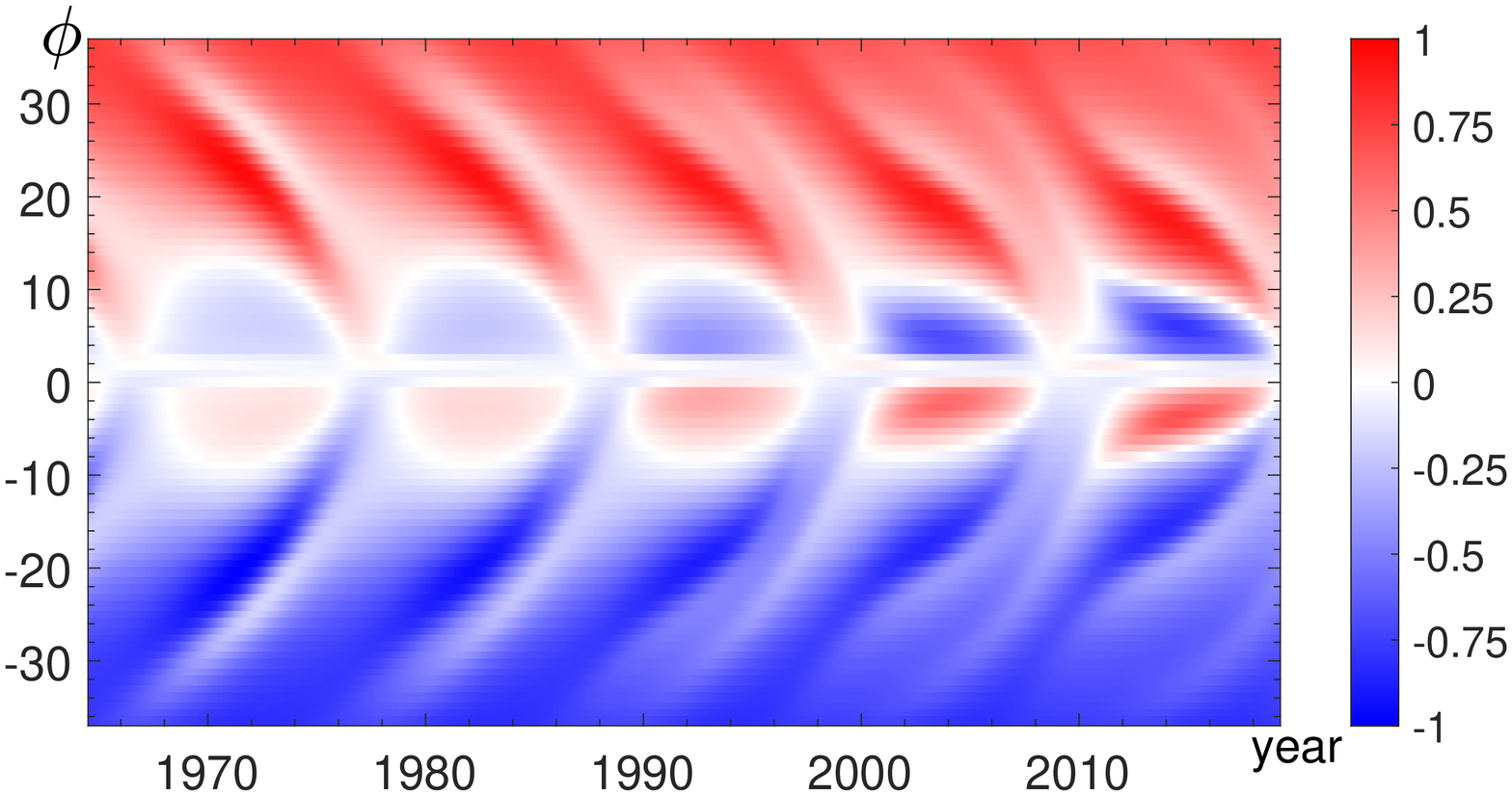}
\includegraphics[width=8.5cm]{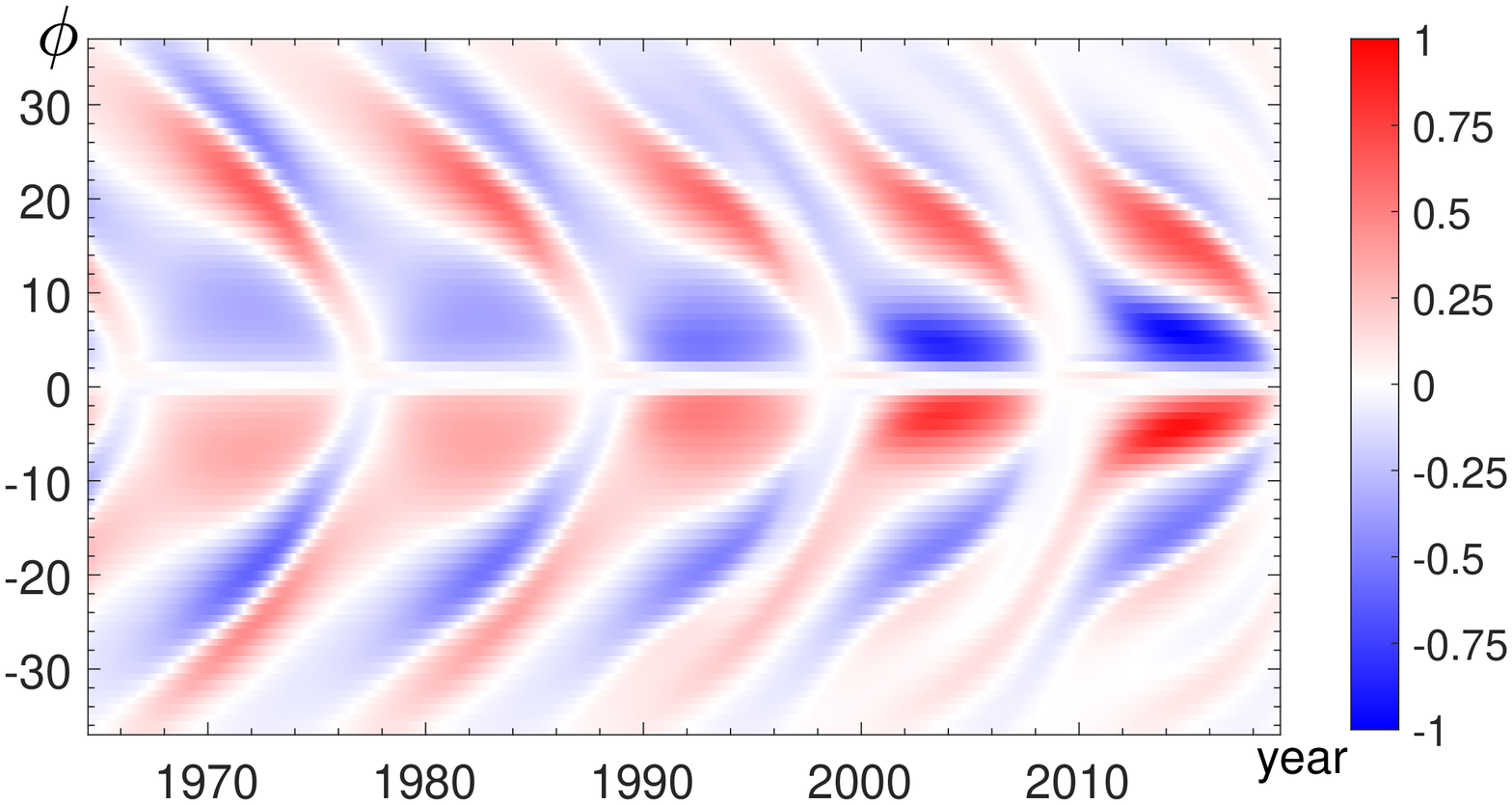}
\includegraphics[width=8.5cm]{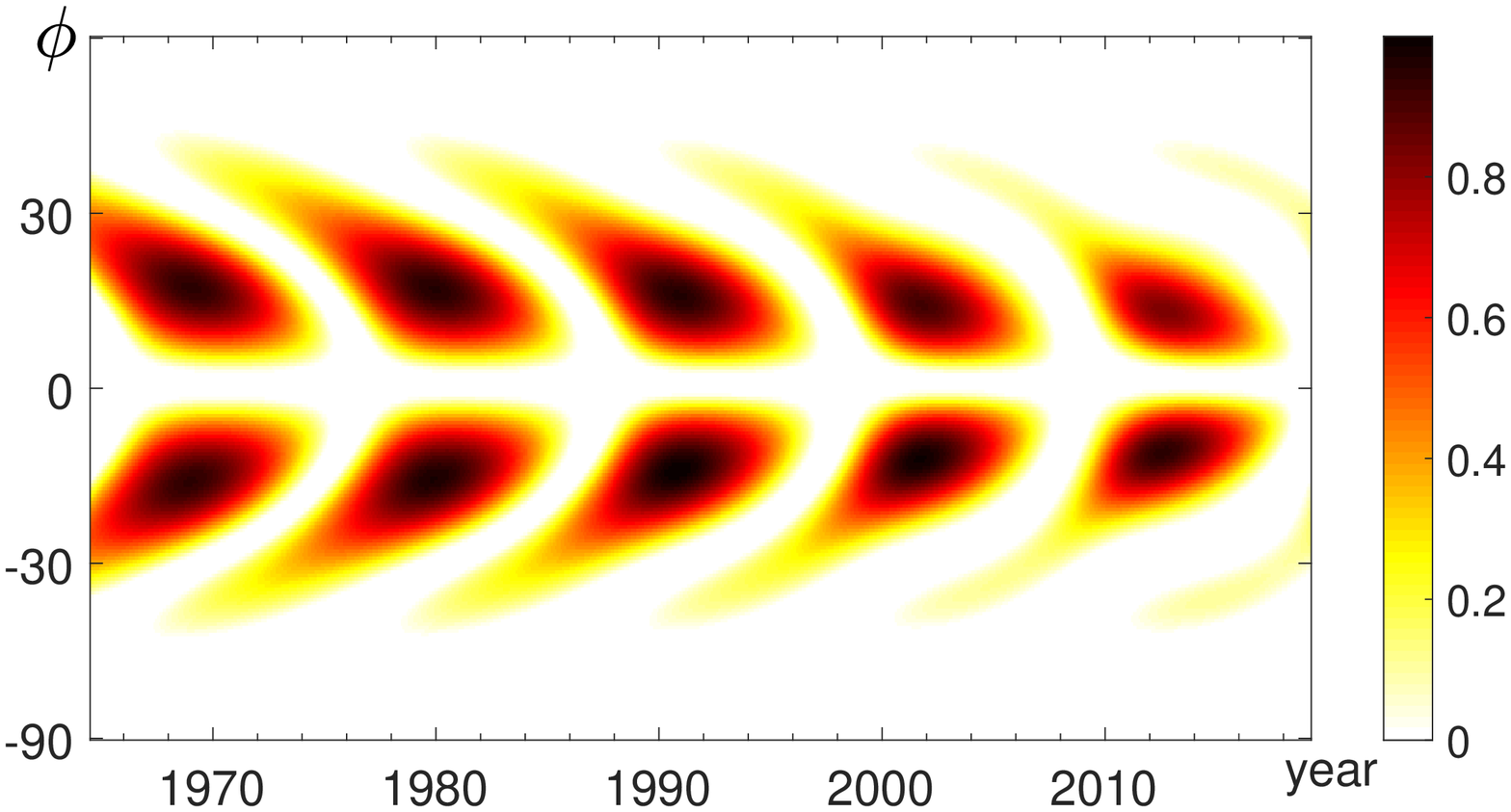}
\caption{\label{Fig3}
Butterfly diagrams of the normalised mean tilt $-\gamma / \tilde\delta_0$ (upper panel)
given by equation~(\ref{RQ2})
and the total magnetic contribution $\delta_{_{M}}^\ast$ to the mean tilt (middle panel)
given by the second line of equation~(\ref{RQ2}).
Butterfly diagram of the surface density of the Wolf numbers
(bottom panel). Here the colour bars are normalised by their maximum values.
}
\end{figure}

Using results of these numerical simulations, we plot in
Fig.~\ref{Fig3} (upper panel) the butterfly diagram
of the normalised mean tilt $-\gamma / \tilde\delta_0$
given by equation~(\ref{RQ2}) with the magnetic contribution to the mean tilt as
\begin{eqnarray}
\delta_{_{M}}\left(\overline{\bm B}^2\right) &=& C_\ast  \, \biggl({\partial^3  \over \partial \phi^3}\, {\overline{\bm B}^2 \over B_{\rm eqp}^2} \biggr)_{\rm bot} ,
\label{B31}
\end{eqnarray}
where the parameter $C_\ast = 0.8$.
The increase of the values of the mean tilt in the recent three cycles in the low latitudes seen in Fig.~\ref{Fig3} (upper panel) can be explained by the joint effect of dipole and quadrupole dynamo modes.
In particular, as follows from the numerical simulations during the nonlinear evolution caused by the dynamics of the magnetic helicity in the recent three cycles, the contribution of the dipole dynamo mode to magnetic activity decreases while the quadrupole dynamo mode contribution increases. This is in a qualitative agreement with observations \citep{LO06}.
In addition, as follows from observations, during the transition from high to low solar cycles, the magnitude
of the mean tilt decreases \citep{DS10}.

In Fig.~\ref{Fig3} (middle panel) we also show the butterfly diagram of the total magnetic contribution $\delta_{_{M}}^\ast = - \tilde \delta_{_{M}} \, (\cos \phi + \tilde \delta_3 \, \cos 3\phi - \tilde \delta_5 \, \cos 5\phi)$ to the mean tilt where the latitudinal part of the differential rotation is taking into account.
For comparison, in Fig.~\ref{Fig3} (bottom panel) we also plot the butterfly diagram of the surface density of the Wolf numbers.
The butterfly diagram of the normalised mean tilt of sunspot bipolar regions is essentially different from that of the surface density of the Wolf numbers. In particular, the mean tilt distribution in every hemisphere is nearly homogeneous, i.e., it depends weakly on the phase of the solar cycle except for small regions for the lower latitudes where the mean tilt has opposite signs in every hemisphere. On the other hand, the distribution of the surface density of the Wolf number is strongly inhomogeneous, i.e., it strongly depends on the phase of the solar cycle.

One can see from Fig.~\ref{Fig3} (middle panel) that around the solar maximum in middle latitudes (the ''Royal" activity zone) the magnetic contribution $\delta_{_{M}}^\ast$ to the mean tilt has the same sign to that of the contribution caused by the Coriolis force. The ''Royal" activity zone migrates towards lower latitudes for lower solar activity circles 23 -- 24, see Fig.~\ref{Fig3} (bottom panel).
In lower latitudes (below $10^\circ$) the magnetic contribution $\delta_{_{M}}^\ast$ to the mean tilt is negative/positive in the north/south hemisphere, see Fig.~\ref{Fig3} (middle panel). This effect increases towards lower solar activity circles 23 -- 24.
In spite of the fact that the magnetic contribution $\delta_{_{M}}^\ast$ in lower latitudes
is the dominant contribution to the mean tilt of sunspot
bipolar regions, its contribution to the mean tilt
is also important at latitudes around $25^\circ$--$30^\circ$
[see Figs.~\ref{Fig1}--\ref{Fig2} and Fig.~\ref{Fig3} (upper and middle panels)].

\section{Comparison with observations of the mean tilt}

In this section we compare our numerical results with observational data of the mean tilt $-\gamma$ of sunspot bipolar regions.
We use the observational data which have been obtained by \cite{TT18} from daily sunspot drawings taken at Mount Wilson Observatory (MWO). The data cover a century long period.
The original MWO drawings were digitized using software package developed by \cite{TVP15}, see also references therein. The digitization includes the date and time of observations, heliographic coordinates of each umbra, its area, the strength, and polarity of its magnetic field. The overall digitized dataset used by us contains 20,318 days of observations from 1917 to October 2016. The method of \cite{TT18} enables to identify clusters of sunspots of positive and negative polarity, from which bipolar pairs have been formed.

There has been the total of 441,973 measurements of the magnetic field of individual nuclei and pores of sunspots carried out, and the total number of 51,413 bipolar regions allocated. Initially, clusters of active regions of positive and negative polarity were searched for. For achieving this, the sunspots were sorted by area for each day of observation, and kernels of the same polarity located at a distance of no more than 10 degrees in longitude and 7 degrees in latitude from the spot of maximum area were selected. For each cluster, the average coordinates were found, which were computed using the weight function over the area. Next, a bipole counterpart through clusters of sunspot negative polarity was found.

The observational data are two-fold. The first group of the data used in the present study to produce Figures~\ref{Fig1} and~\ref{Fig2} (see Section~2), is the result of averaging of bipolar pairs of all sizes. This group of the data is presented in Tables~1 and~2 in \cite{TT18}, where the mean value and the standard deviation of Gaussian fittings have been computed. We use the data to compare with the mean tilts obtained from our theoretical and numerical simulations. We have shown that the theoretical results fit the observations very well.
The data have been filtered out by the bipolar regions smaller in length than 3 degrees. In total there were 18,547 bipolar regions in the even and 17,435 in the odd solar cycles. We used the bipolar regions greater than 3 degrees because smaller bipolar regions almost do not possess a certain tilt angles.

\begin{figure}
\centering
\includegraphics[width=8.5cm]{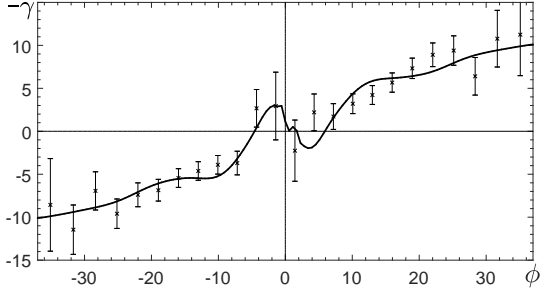}
\caption{\label{Fig4}
The mean tilt $-\gamma$ (in degrees) versus the latitude $\phi$ (in degrees): numerical simulations (solid line)
and observations of sunspot bipolar regions (dashed line) averaged over individual cycles 15 - 24.}
\end{figure}

The second group of the data used below to produce Figure~\ref{Fig4} is comprised of the all data on the tilts of all bipolar regions filtered by the small sized bipolar pairs, so that only the bipolar regions larger by size than 3 degrees were retained. The cut-off area of those pairs was set to several $\mu$H ($4\pi \times 10^{-6}$ of steradian).
We have used those data to confront with our theoretical and numerical results based
on the time evolution of the mean tilt of sunspot bipolar regions.

Note that both these samplings are very different from that was earlier published for statistics of bipolar regions by \cite{TIS13,T15}. In earlier works the bipolar regions have been composed from individual sunspot nuclei, while in our studies they are formed from the clusters of sunspots. Thus, our results may be qualitatively very different from those of \cite{KS08,DS10}. Since the nuclei of spots are formed of the two opposite polarities, the technique and the results are significantly different.

In Figure~\ref{Fig4} we show the mean tilt $-\gamma$ (in degrees) versus the latitude $\phi$ (solid line) obtained using equation~(\ref{RQ2}), where the magnetic contribution $\delta_{_{M}}$ to the mean tilt is calculated by the mean-field numerical simulation for $C_\ast = 0.8$, $\delta_0 =0.29$, $\delta_3 = 0.122$,
$\delta_5 =  1.56 \times 10^{-2}$, $\tilde \delta_3 = 4.48$, and $\tilde \delta_5 = 1.02$.
These numerical results are also compared with the observational data of the mean tilt $-\gamma$ of sunspot bipolar regions.
The observational data have been averaged over individual solar cycles (from the cycle 15 to 24).
The numerical results are also averaged over the same cycles.
It follows from Figure~\ref{Fig4} that there is an asymmetry between the northern and southern hemisphere.
We stress that we have taken into account here an effect of the latitudinal dependence of the solar differential rotation on the mean tilt of the sunspot bipolar regions as well as the contribution to the mean tilt
caused by the large-scale magnetic field.
The obtained theoretical results and performed numerical simulations
for the mean tilt of sunspot bipolar regions are in an agreement with
the observational data.

Remarkably, that there is a difference between Figure~\ref{Fig1} and Figure~\ref{Fig4} in the vicinity of the equator.
In particular, in Figure~\ref{Fig4} the mean tilt is calculated by averaging over only large-size bipolar regions, and it is not zero in the vicinity of the equator. Moreover, the mean tilt of the large-size bipolar regions
reverses its sign in the vicinity of the equator $\phi \approx \pm 5^\circ$.
On the other hand, in Figure~\ref{Fig1} the mean tilt is calculated by averaging over active regions of all sizes, and it tends to zero in the vicinity of the equator.
The explanation of this fact is given in the next section.

\section{The contribution of the current helicity to the mean tilt}
\label{Sect-4}

The current helicity, $\langle {\bm B}^{\rm ar} {\bf \cdot} \bec{\rm curl} \,{\bm B}^{\rm ar} \rangle$,
of the magnetic field, ${\bm B}^{\rm ar}$, in the active region describes the correlation between
the magnetic field and the electric current, and it characterises the twist of the magnetic field,
where the angular brackets denote averaging over the surface occupied by the active region.
This implies that the current helicity of the active region should contribute to the total mean tilt.
This contribution is given by
\begin{eqnarray}
\gamma_{_{H}} = L_{\rm ar} {\langle {\bm B}^{\rm ar} {\bf \cdot} \bec{\rm curl} \,{\bm B}^{\rm ar} \rangle
\over \langle \left({\bf B}^{\rm ar}\right)^2\rangle} ,
\label{H1}
\end{eqnarray}
where $L_{\rm ar}$ is the characteristic size of an active region.
It has been shown by \cite{ZKR12}, that the
current helicity of the active region,
$\langle {\bm B}^{\rm ar} {\bf \cdot} \bec{\rm curl} \,{\bm B}^{\rm ar} \rangle$, is related to the magnetic helicity $\langle {\bm A}^{\rm ar}{\bf \cdot}{\bm B}^{\rm ar}\rangle$ of the active region as
\begin{eqnarray}
\langle {\bm B}^{\rm ar} {\bf \cdot} \bec{\rm curl} \,{\bm B}^{\rm ar} \rangle \approx {1\over L^2_{\rm ar}} \, \langle {\bm A}^{\rm ar}{\bf \cdot}{\bm B}^{\rm ar} \rangle + O\biggl({L^2_{\rm ar} \over R_\odot^2}\biggr) ,
\label{H2}
\end{eqnarray}
where $R_\odot$ is the solar radius.
Substituting equation~(\ref{H2}) to equation~(\ref{H1}), we obtain
\begin{eqnarray}
\gamma_{_{H}} = {\langle {\bm A}^{\rm ar}{\bf \cdot}{\bm B}^{\rm ar} \rangle
\over L_{\rm ar} \, \langle \left({\bf B}^{\rm ar}\right)^2\rangle} .
\label{H3}
\end{eqnarray}
The total magnetic helicity $H_{\rm total}$ is conserved.
Due to a non-zero flux of magnetic helicity, a part of the total magnetic helicity
is transported to chromosphere and corona from the active region.
This implies that the total magnetic helicity
$H_{\rm total}\equiv \langle {\bm A}^{\rm ar}{\bf \cdot}{\bm B}^{\rm ar} \rangle \, V$ is the sum
of the transported magnetic helicity, $H_{\rm transp}$, and residual magnetic helicity (i.e., observable magnetic helicity), $H_{\rm observ}$:
\begin{eqnarray}
H_{\rm total}= H_{\rm transp} +  H_{\rm observ} ,
\label{H4}
\end{eqnarray}
where $V$ is the volume occupied by the active region.
Here we assume that the transported magnetic helicity, $H_{\rm transp}$, is a sum
of the magnetic helicity caused by writhe of the active region,
$C_w \langle {\bm A}^{\rm ar}{\bf \cdot}{\bm B}^{\rm ar} \rangle \, V$ (with $C_w < 1)$,
and the produced magnetic helicity, $C_\Omega  \, \langle \left({\bf B}^{\rm ar}\right)^2\rangle \, \gamma \, L_{\rm ar} \, V$ by mechanical twist of magnetic flux tube due to the Coriolis force:
\begin{eqnarray}
H_{\rm transp} = \left(C_w \, \langle {\bm A}^{\rm ar}{\bf \cdot}{\bm B}^{\rm ar} \rangle +
C_\Omega  \, \langle \left({\bf B}^{\rm ar}\right)^2\rangle \, \gamma \, L_{\rm ar} \right) V ,
\label{H5}
\end{eqnarray}
where $\gamma$ is the mean tilt discussed in Sects.~2-4.
The observed mean tilt is defined as
\begin{eqnarray}
\gamma_{\rm observ} = {H_{\rm observ} \over \langle \left({\bf B}^{\rm ar}\right)^2\rangle \, L_{\rm ar} \, V} .
\label{H6}
\end{eqnarray}
The total magnetic helicity $H_{\rm total} \equiv \gamma_{_{H}} L_{\rm ar} \, \langle \left({\bf B}^{\rm ar}\right)^2\rangle \,  V$ is conserved.
Substituting equations~(\ref{H5}) to equation~(\ref{H4}), and using equations~(\ref{H3}) and~(\ref{H6}), we obtain
\begin{eqnarray}
\gamma_{\rm observ} = (1- C_w) \, \gamma_{_{H}} - C_\Omega \, \gamma .
\label{H7}
\end{eqnarray}
Since $\gamma_{_{H}} \propto L_{\rm ar}^{-1}$ and $\gamma \propto L_{\rm ar}^{2}$, we obtain that $|\gamma_{_{H}}| \ll |\gamma|$ for large-size active regions, where we take into account that $\delta_0 \propto L_B^2 \geq L_{\rm ar}^2$.
This implies that $C_\Omega=-1$, because in this case $\gamma_{\rm observ} \approx \gamma$.
On the other hand, for small-size active regions $|\gamma_{_{H}}| \gg |\gamma|$.
In general case, the both terms in equation~(\ref{H7}) are important, so that the observable tilt is given by
\begin{eqnarray}
\gamma_{\rm observ} = \gamma + (1- C_w) \, {\langle {\bm A}^{\rm ar}{\bf \cdot}{\bm B}^{\rm ar} \rangle
\over L_{\rm ar} \, \langle \left({\bf B}^{\rm ar}\right)^2\rangle} ,
\label{H8}
\end{eqnarray}
where we use equation~(\ref{H1}).
It follows from this equation that there is a size of bipolar region where  both contributions
to the mean tilt are of the same order.
This is in a qualitative agreement with study by \cite{ITS15}.
Note that $\langle {\bm A}^{\rm ar}{\bf \cdot}{\bm B}^{\rm ar} \rangle=
- \meanAA {\bf \cdot} \meanBB$ \citep{ZKR12}, so that
\begin{eqnarray}
\gamma_{\rm observ} = \gamma - (1- C_w) \, {\meanAA {\bf \cdot} \meanBB
\over L_{\rm ar} \, \langle \left({\bf B}^{\rm ar}\right)^2\rangle} .
\label{HH8}
\end{eqnarray}
Using equations~(\ref{RQ2}) and~(\ref{HH8}), we obtain
\begin{eqnarray}
\gamma_{\rm observ} &=& - \tilde \delta_0 \Big[\sin \phi + \delta_3 \sin 3\phi - \delta_5 \sin 5\phi
\nonumber\\
&&- \tilde \delta_{_{M}} \, \Big(\cos \phi + \tilde \delta_3 \, \cos 3\phi - \tilde \delta_5 \, \cos 5\phi\Big) \Big]
\nonumber\\
&& - (1- C_w) \, {\meanAA {\bf \cdot} \meanBB
\over L_{\rm ar} \, \langle \left({\bf B}^{\rm ar}\right)^2\rangle} .
\label{H9}
\end{eqnarray}
In the north hemisphere, $\meanAA {\bf \cdot} \meanBB$ is negative,
which implies that the sign of the last term describing the contribution caused by
the magnetic helicity of active region, is opposite to that due
to the Coriolis force contribution (the $\gamma$ term).
This implies that the mean tilts caused by the large-size and small-size active regions have
opposite signs.

This explains the difference between Figure~\ref{Fig1} and Figure~\ref{Fig4} in the vicinity of the equator.
Indeed, in Figure~\ref{Fig4} the mean tilt determined by averaging over only large-size bipolar regions, is not zero in the vicinity of the equator. Contrary, in Figure~\ref{Fig1} the mean tilt determined by averaging over all active regions, tends to zero in the vicinity of the equator.
The physical reason for this fact is as follows.
Since the mean tilt caused by the large-size and small-size active regions have
opposite signs, the mean tilt calculated by averaging over all active regions is small,
because their contributions compensate each other.
This implies that in the vicinity of equator the mean tilt
is less than that calculated by averaging over only large-size bipolar regions.
We remind that the effect of Coriolis force on the mean tilt vanishes in the vicinity of equator,
so that the above effect of the compensation of the contributions to the mean tilt caused by the large-size and small-size active regions is more pronounced in the vicinity of equator.

%\medskip

\section{Conclusions}

We have developed a theory of the mean
tilt of sunspot bipolar regions.
The formation of the mean tilt is caused by the
effect of the Coriolis force on meso-scale motions of super-granular convection
and large-scale meridional circulation.
We have demonstrated that at low latitudes
the joint action of the Coriolis force and the magnetic tension
results in an additional magnetic contribution to the mean tilt of the sunspot bipolar regions
which depends on the large-scale magnetic field.
We have also found an additional contribution to the mean tilt of the sunspot bipolar regions
which is caused by an effect of the latitudinal dependence of the solar differential rotation
on the mean tilt.
The latter can explain the deviations from the Joy's law for the mean tilt at higher latitudes.
The obtained theoretical results and performed numerical simulations
for the mean tilt are in an agreement with
the observational data of the mean tilt of the sunspot bipolar regions.

\section*{Acknowledgments}

The detailed comments on our manuscript by the anonymous referee
are very much appreciated.
Stimulating discussions with participants
of the NORDITA programs on ''Solar Helicities in Theory and Observations:
Implications for Space Weather and Dynamo Theory"
(March 2019) and ''The shifting paradigm of stellar convection:
from mixing length concept to realistic turbulence modelling"
(March 2020) are acknowledged.
The work of KK and NS was supported in part
by a grant from the Russian Science Foundation
(grant RNF 18-12-00131) at the Crimean Astrophysical Observatory.
AT would like to acknowledge support from RFBR grant 18-02-00098
for the observational data analysis.
NK, KK, IR acknowledge the hospitality of NORDITA.

\appendix

\section{\bf Derivation of equations~(\ref{B5}), (\ref{B14}) and~(\ref{RQ2})}
\label{Appendix-A}

To derive equation~(\ref{B5}), we rewrite equations~(\ref{B2}) and~(\ref{B3}) for small perturbations $\tilde{\bm b}$ and $\tilde{s}$ as
\begin{eqnarray}
\tilde{\bm b} = ({\bm B}^{\rm eq} \cdot {\bm \nabla}) {\bm \xi} - ({\bm \xi} \cdot {\bm \nabla}) {\bm B}^{\rm eq} - \Lambda_\rho \, {\bm B}^{\rm eq} \, (\hat{\bm r}  \cdot {\bm \xi})  ,
\label{BBB2}
\end{eqnarray}
\begin{eqnarray}
\tilde{s} = - ({\bm \xi} \cdot {\bm \nabla}) {S}^{\rm eq} - {\Omega_b^2 \over g} \, {\bm \xi} \cdot \hat{\bm r} .
 \label{BBB3}
\end{eqnarray}
Substituting equations~(\ref{BBB2}) and~(\ref{BBB3}) into equation~(\ref{B1})
rewritten for small perturbations ${\bm \xi}$, we obtain equation~(\ref{B5}). In derivation of equation~(\ref{B5}), we use assumptions~(\ref{AB5}) outlined in Section~2.

To derive equation~(\ref{B14}) for the mean tilt of sunspot bipolar regions, we exclude the pressure term from equation~(\ref{B5}) by applying {\bf curl} to this equation and multiply the obtained
equation by unit vector ${\bm e}_B={\bm B}^{\rm eq} / {B}^{\rm eq}$.
This yields
\begin{eqnarray}
&& {\partial^2 \tilde \gamma \over \partial t^2}  = 2 \left[{\bm \nabla} \times \left(\left({\bm U}^{\rm eq} + {\partial {\bm \xi} \over \partial t} + {\bm v}^{({\rm c})} \right) \times {\bm \Omega}\right)\right] \cdot {\bm e}_B
\nonumber\\
 &&\quad + \left({\bm U}_{\rm A} \cdot {\bm \nabla}\right)^2
\delta_B ,
 \label{B7}
\end{eqnarray}
where $\tilde \gamma= {\bm \delta}^{\rm tw} \cdot {\bm e}_B$ is the tilt, ${\bm \delta}^{\rm tw} = {\bm \nabla} \times {\bm \xi}$.
Here we take into account that at the boundary between the convective zone and the photosphere, the magnetic field inside the sunspots is preferably directed in the radial direction.
Since the second and the last terms in equation~(\ref{B5}) are directed in the radial direction,
they do not contribute to the $\hat{\bm r}$ component of the {\bf curl}, i.e., they do not contribute to $\tilde \gamma$.

We seek for the solution of equation~(\ref{B7}) in the form of standing Alfv\'{e}n waves as
\begin{eqnarray}
\tilde \gamma = \sum_{m=0}^\infty A_m  \cos \left[{(2m + 1) \pi \zeta \over L_B}\right]  \cos \left[{2 \pi \, t \over T_m} + \varphi \right] ,
%\nonumber\\
 \label{B8}
\end{eqnarray}
where $T_m=2 \tau_{_{\rm A}} / (2m + 1)$ is the period of non-dissipating oscillations,
$\tau_{_{\rm A}} = L_B /{U}_{\rm A}$ is the  Alfv\'{e}n time, $\zeta$ is the coordinate
along the magnetic field line of the length $L_B$ connecting sunspots of the opposite magnetic polarities. Now we take into account that
$T_m \, \Omega \ll 1$, which implies that $|\partial {\bm \xi} / \partial t|
\ll |{\bm v}^{({\rm c})}|, |{\bm U}^{\rm eq}|$.
We also assume that the source of the tilt $I_\gamma=2 \Big[{\bm \nabla} \times [({\bm U}^{\rm eq} + {\bm v}^{({\rm c})}) \times {\bm \Omega}]\Big] \cdot {\bm e}_B$ in equation~(\ref{B7}) is localized
near the boundary between the solar convective zone and the photosphere.
This source can be modelled as the combination of two Dirac delta-functions:
\begin{eqnarray}
I_\gamma(\zeta)&=&2 \Big[{\bm \nabla} \times [({\bm U}^{\rm eq} + {\bm v}^{({\rm c})}) \times {\bm \Omega}]\Big] \cdot {\bm e}_B
\nonumber\\
&&\times \Big[\delta(\zeta/L_B) - \delta(\zeta/L_B -1)\Big] ,
\label{B9}
\end{eqnarray}
where $\delta(x)$ is the Dirac delta-function.

We substitute equation~(\ref{B8}) into equation~(\ref{B7}) and after the Fourier transformation of the source term~(\ref{B9}), we obtain equation for the amplitude $A_m(t)$ as
\begin{eqnarray}
{\partial^2 A_m \over \partial t^2}  = {2 I_\gamma \over \pi} - \left[{U}_{\rm A} \,  {(2m + 1) \pi \over L_B}\right]^2 A_m .
\label{B10}
\end{eqnarray}
This equation with initial condition $A_m(t=0)=0$ has the following solution:
\begin{eqnarray}
A_m(t) = {2 I_\gamma \, \tau_{_{\rm A}}^2 \over \pi^3(2m + 1)^2} \, \left\{1 - \cos \left[{(2m + 1) \, \pi \, t \over \tau_{_{\rm A}}}\right]\right\} .
\nonumber\\
\label{B11}
\end{eqnarray}
Substituting equation~(\ref{B11}) into equation~(\ref{B8}), we obtain expression for $\tilde \gamma$ as
\begin{eqnarray}
\tilde \gamma &=& {2 I_\gamma \, \tau_{_{\rm A}}^2 \over \pi^3} \, \sum_{m=0}^\infty {1 \over (2m + 1)^2} \,  \cos \left[{(2m + 1) \pi \zeta \over L_B}\right]
\nonumber\\
&& \quad \quad \quad \quad\quad\times \left\{1 - \cos \left[{(2m + 1) \, \pi \, t \over \tau_{_{\rm A}}}\right]\right\} .
 \label{B12}
\end{eqnarray}
Averaging equation~(\ref{B12}) over the time that is larger than the maximum Alfv\'{e}n time $\tau_{_{\rm A}}$, we obtain equation~(\ref{B14}) for the mean tilt $\gamma=\langle\tilde \gamma\rangle_{\rm time}$ of sunspot bipolar regions at the surface of the sun.

For the derivation of equation~(\ref{RQ2}) we used identities given below:
\begin{eqnarray}
&& \sin^3\phi = {1 \over 4} \left[3 \sin \phi - \sin 3\phi\right] ,
\label{RP1}
\end{eqnarray}
\begin{eqnarray}
&& \sin^5\phi = {1 \over 16} \left[10 \sin \phi - 5 \sin 3\phi + \sin 5\phi\right] ,
\label{RP2}
\end{eqnarray}
\begin{eqnarray}
&& \sin^2\phi \, \cos \phi = {1 \over 4} \left[\cos \phi - \cos 3\phi\right] ,
\label{RP3}
\end{eqnarray}
\begin{eqnarray}
&& \sin^4\phi \, \cos \phi = {1 \over 16} \left[2 \cos \phi - 3 \cos 3\phi + \cos 5\phi\right] .
\label{RP4}
\end{eqnarray}

\section{Equation for the radial mean velocity}
\label{Appendix-B}

The momentum equation~(\ref{B1}) with additional force caused by the eddy
viscosity in a steady state in spherical coordinates reads:
\begin{eqnarray}
{\partial \over \partial r} \overline{P}_{\rm tot} &=& {2 \over r^2} {\partial \over \partial r}
\biggl(r^2 {\rho_0 \, \overline{U}_r \, \nu_{_{T}} \over H_\rho } \biggr)
- {\overline{B}_{\varphi}^{2} \over 4 \pi r}
+ 2 \rho_0 \, \overline{U}_\varphi \, \Omega \, \sin \theta
\nonumber\\
&& + {1 \over r \, \sin \theta} {\partial \over \partial \theta} \biggl(\sin \theta \,
{\rho_0 \, \overline{U}_\theta \, \nu_{_{T}} \over H_\rho}\biggr) ,
\label{R1} \\
\end{eqnarray}
\begin{eqnarray}
{\partial \over \partial \theta } \overline{P}_{\rm tot} &=&  {1 \over r^2} {\partial \over \partial r}
\biggl( r^3 {\rho_0 \, \overline{U}_\theta \, \nu_{_{T}}  \over H_\rho} \biggr)
- {\overline{B}_{\varphi }^{2} \over 4 \pi } \cot \theta
\nonumber\\
&& + 2 \rho_0 \, \overline{U}_\varphi \, \Omega \, r \, \cos \theta ,
\label{R2}
\end{eqnarray}
Here $\overline{P}_{\rm tot}=\overline{P} + \overline{\bm B}^2/ 8 \pi + \overline{\bm U}^2 / 2$ is the total pressure,
$H_\rho$ is the density height scale, and $\nu_{_{T}}$ is
the eddy viscosity.

We exclude the total pressure term, use the continuity equation
${\bm \nabla} \cdot (\rho_0 \, \overline{\bm U}) = 0$, and introduce the stream function
$\Psi$:
\begin{eqnarray}
\rho_0 \, \overline{U}_r = {1 \over r^2 \sin \theta}  {\partial \Psi \over \partial
\theta} , \quad \rho_0 \, \overline{U}_\theta = - {1 \over r \sin \theta} {\partial
\Psi \over \partial r} .
\label{R8}
\end{eqnarray}
After neglecting a week dependence of $\nu_{_{T}} /
H_\rho$ on radius $r$, equations~(\ref{R1})--(\ref{R2}) are reduced to
\begin{eqnarray}
{\partial^2 Y \over \partial X^2} + {1 \over 9 X^2}
{\partial \over \partial \theta} \biggl( {1 \over \sin
\theta} {\partial \over \partial \theta} (Y \sin \theta) \biggr) =
f(X,\theta) ,
\label{R3}
\end{eqnarray}
where $X= r^3 , \quad Y = X \, \rho_0 \, \overline{U}_\theta \, \nu_{_{T}} / H_\rho$, and
\begin{eqnarray}
f(X,\theta) = {1 \over 36 \pi } \biggl( {1 \over X} {\partial
\over \partial \theta}  - {3 \over \tan \theta}
{\partial \over \partial X} \biggr)\overline{B}^2_\varphi .
\label{R9}
\end{eqnarray}
Here we take into account that the contribution of the Coriolis
force into the function $f(X,\theta)$ under condition of the slow
rotation is small \citep{KR91,KMR96}. The solution of
equation~(\ref{R3}) with the boundary condition
\begin{eqnarray}
\left[(1 - \kappa) \, {\partial (\rho_0 \, \overline{U}_r) \over \partial r}
+ {2 \rho_0 \, \overline{U}_r \over r}\right]_{r=R_\odot} = 0,
\label{R4}
\end{eqnarray}
is given by
\begin{eqnarray}
\overline{U}_r = {\ell_0^2 \over 4 \pi \, \kappa \, \nu_{_{T}} \, \rho_{\rm top} \, R_\odot} \, {1 \over \sin \theta}\, {\partial  \over \partial \theta} \Big[\sin \theta F(\theta)\Big] ,
 \label{R5}
\end{eqnarray}
where the parameter $\kappa \approx 0.3$ -- $0.4$ characterises a fraction of the large-scale
radial momentum of plasma  which is lost
during crossing the boundary between the convective zone and photosphere, and
\begin{eqnarray}
F(\theta) \approx \int_{R_\odot - L}^{R_\odot} \left(1 + {R_\odot - r \over L - \ell_0}\right) \,
\left({\partial \overline{\bm B}^2 \over \partial \theta} \right) {\,d r \over r}
\nonumber\\
\approx C_u \, \left({\partial \overline{\bm B}^2 \over \partial \theta} \right)_{bot} ,
 \label{R6}
\end{eqnarray}
where the constant $C_u$ varies from $0.7$ to $1$ depending on the radial profile of the mean magnetic field.
Therefore, equations~(\ref{R5})--(\ref{R6}) yield equation~(\ref{R7}).

\section{The evolution of the Wolf number}
\label{Appendix-C}

In the framework of the nonlinear mean-field dynamo model by
Kleeorin et al. (2016) and Safiullin et al. (2018),
the phenomenological budget equation for the surface density of the Wolf number is given by
\begin{eqnarray}
{\partial \tilde W \over \partial t} = I_{\rm w}(t,\theta) - {\tilde W \over \tau_s(\meanB)} ,
\label{TM12}
\end{eqnarray}
where the rate of production of the surface density of the Wolf number caused
by the formation of sunspots is
\begin{eqnarray}
I_{\rm w}(t,\theta) = {|\gamma_{\rm inst}| |\meanB-\meanB_{\rm cr}| \over \Phi_s}
\Theta(\meanB-\meanB_{\rm cr}) ,
\label{TM14}
\end{eqnarray}
and the rate of decay of sunspots is $\tilde W / \tau_s(\meanB)$
with the decay time, $\tau_s(\meanB)$, of sunspots
and $\Theta(x)$ is the $\Theta$ function, defined
as $\Theta(x) = 1$ for $x>0$, and $\Theta(x) = 0$ for $x\leq 0$.
Here $\meanB_{\rm cr}$ is the threshold for the sunspot formation
and $\gamma_{\rm inst}$ is the inverse time of
the formation of sunspots.

As an example for estimation of the parameters $\overline{B}_{\rm cr}$ and $\gamma_{\rm inst}$,
we use in the present study the negative effective magnetic pressure instability
\citep{KRR89,KRR90,KMR93,KMR96,KR94,RK07}
resulting in formation of magnetic spots \citep{BKR11,BKR13,BGKR14}
and bipolar active regions \citep{WKR13,WKR16}.
The growth rate $\gamma_{\rm inst}$ of the negative effective magnetic pressure instability
is given by
\begin{eqnarray}
\gamma_{\rm inst} &=&  \left( {2 \overline{U}_{\rm A}^2 k_x^2 \over H_\rho^2 k^2}
\left|{d P_{\rm eff} \over d \beta^2}\right| - {4 ({\bm \Omega} \cdot {\bm k})^2
\over {\bm k}^2}\right)^{1/2}
\nonumber\\
&& - \eta_{_{T}} \left(k^2 + {1 \over (2 H_\rho)^{2}}\right),
\label{TM15}
\end{eqnarray}
\citep{RK07,BRK16}, where $\overline{U}_{\rm A}=\overline{\rm B}/ (4 \pi \rho_0)^{1/2}$ is the Alfv\'{e}n speed based on the mean magnetic field,
${\bm k}$ is the wave number, $P_{\rm eff}=\half\left[1-q_{\rm p}(\beta)\right]\beta^2$ is the effective magnetic pressure, the nonlinear function $q_{\rm p}(\beta)$ is the turbulence contribution to the mean magnetic pressure and $\beta=\meanB/\meanB_{\rm eqp}$.
We assume here that the characteristic time of the Wolf number variations is of the order of the characteristic time for excitation of the instability, $\gamma_{\rm inst}^{-1}$.
When the instability is not excited ($\gamma_{\rm inst}<0$), the production rate of sunspots, $I_{\rm w}(t,\theta) \to 0$, which means that the function $I_{\rm w}(t,\theta) \propto |\gamma_{\rm inst}| \, \Theta(\meanB-\meanB_{\rm cr})$.
The production term  of sunspots is also proportional to the maximum number of sunspots per unit area, which is estimated as $\sim |\meanB-\meanB_{\rm cr}| /\Phi_s$, where $|\meanB-\meanB_{\rm cr}|$ is the magnetic flux per unit area that contributes to the sunspot formation and $\Phi_s$ is the magnetic flux inside a magnetic spot.
This instability is excited when the mean magnetic field is larger than a critical value,
$\meanB>\meanB_{\rm cr}$:
\begin{eqnarray}
{\meanB_{\rm cr} \over \meanB_{\rm eq}} = {\ell_0 \over 50 H_\rho} \left[1 + \left({10 \, {\rm Co} \, H_\rho^2 \over \ell_0^2} \right)^2 \right]^{1/2} .
\label{TM20}
\end{eqnarray}
This instability is excited in the upper part of the convective zone,
where the Coriolis number ${\rm Co}= 2 \Omega \, \tau$ is small.
The decay time $\tau_s(\meanB)$ varies from several weeks to a couple of month, while the solar cycle period is about 11 years. This allows us use the steady-state solution of Eq.~(\ref{TM12}), $\tilde W = \tau_s(\meanB) \,I_{\rm w}(t,\theta)$. The Wolf number is defined as a surface integral as
$W = R_\odot^2 \, \int \tilde W(t,\theta) \sin \theta \, d\theta \, d\varphi
=2 \pi \,  R_\odot^2 \, \int \tau_s(\meanB) \,I(t,\theta) \sin \theta \,d\theta $.
The function $\tau_s(\meanB)$ is given by
$\tau_s(\meanB)=\tau_\ast \exp \left(C_s \, \partial \meanB/\partial t\right)$,
where $C_s= 1.8 \times 10^{-3}$ and $\gamma_{\rm inst} \, \tau_\ast  \sim 10$.

There are also other mechanisms for the
formation of inhomogeneous magnetic structures, e.g., the magnetic
buoyancy instability (or interchange instability)
of stratified continuous magnetic field
\citep{P66,G70a,P82},
the magnetic flux expulsion \citep{W66}, the
topological magnetic pumping \citep{DY74}, etc.
Magnetic buoyancy applies in the literature for different situations.
The first corresponds to the magnetic buoyancy
instability of stratified continuous magnetic field
\citep{P66,G70a,P82}, and magnetic flux tube concept is not used
there. The second describes buoyancy of discrete magnetic flux tubes
discussed in different contexts in solar physics and
astrophysics \citep{P55,S81,SB82,SC94,DG06,CC07}. This
is also related to the problem of the storage of magnetic
fields in the overshoot layer near the bottom of the solar
convective zone \citep{SW80,T01,TH04}.

The growth rate of the magnetic
buoyancy instability reads
\begin{eqnarray}
\gamma_{\rm inst} &=& {\overline{U}_{\rm A} \over H_\rho} \, \left[Q_{\rm p} \left({H_\rho \over \tilde L_B} -1\right)\right]^{1/2} - \eta_{_{T}} k^2 ,
\label{MB10}
\end{eqnarray}
where $\tilde L_B$ is the characteristic scale of the initial mean magnetic field
variations and $Q_{\rm p}=1-q_{\rm p}(\beta)$.
Without turbulence $Q_{\rm p}=1$ and
the magnetic buoyancy instability of stratified continuous
magnetic field is excited when the scale of variations of the
initial magnetic field is less than the density stratification
length.
The source of a free energy for magnetic buoyancy instability
is the energy of the gravitational field.
Generally, the critical magnetic field $\overline{B}_{\rm cr}$ and the growth rate
$\gamma_{\rm inst}$ for the magnetic buoyancy instability can be used for the estimation
of the rate of production of the Wolf number density given by Eq.~(\ref{TM14}).

However, in the presence of strong turbulence,
$Q_{\rm p}$ can be negative, and
the negative effective magnetic pressure instability can be excited.
The source of a free energy for
the negative effective magnetic pressure instability
is energy of turbulence or turbulent convection.

\end{document}